\documentclass[aps,prl,amsmath,reprint,amssymb,superscriptaddress]{revtex4-2}
\usepackage{epsf}
\usepackage{bm}
\usepackage{hhline}

\usepackage{amsfonts}
\usepackage{latexsym, mathtools}
\usepackage{graphicx}

\newcommand{\e}{$E_{\rm{F}}$}
\newcommand{\m}{$\overline{\rm{M}}$}
\newcommand{\g}{$\overline{\rm \Gamma}$}
\renewcommand{\arraystretch}{0}

\begin{document}

\title{Spin-polarized saddle points in the topological surface states of the elemental Bismuth revealed by a pump-probe spin-resolved ARPES}

\author{Yuto~Fukushima}
\affiliation{Institute for Solid State Physics, The University of Tokyo, Kashiwa, Chiba 277-8581, Japan}

\author{Kaishu~Kawaguchi}
\affiliation{Institute for Solid State Physics, The University of Tokyo, Kashiwa, Chiba 277-8581, Japan}

\author{Kenta~Kuroda}
\affiliation{Graduate School of Advanced Science and Engineering, Hiroshima University, Higashi-hiroshima, Hiroshima 739-8526, Japan}
\affiliation{International Institute for Sustainability with Knotted Chiral Meta Matter (WPI-SCKM$^2$), Hiroshima University, Higashi-hiroshima, Hiroshima 739-8526, Japan}

\author{Masayuki~Ochi}
\affiliation{Department of Physics, Osaka University, Toyonaka, Osaka 560-0043, Japan.}
\affiliation{Forefront Research Center, Osaka University, Toyonaka, Osaka 560-0043, Japan.}

\author{Hiroaki~Tanaka}
\affiliation{Institute for Solid State Physics, The University of Tokyo, Kashiwa, Chiba 277-8581, Japan}

\author{Ayumi~Harasawa}
\affiliation{Institute for Solid State Physics, The University of Tokyo, Kashiwa, Chiba 277-8581, Japan}

\author{Takushi~Iimori}
\affiliation{Institute for Solid State Physics, The University of Tokyo, Kashiwa, Chiba 277-8581, Japan}

\author{Zhigang~Zhao}
\affiliation{Institute for Solid State Physics, The University of Tokyo, Kashiwa, Chiba 277-8581, Japan}
\affiliation{School of Information Science and Engineering, Shandong University, Qingdao, 266237, China}

\author{Shuntaro~Tani}
\affiliation{Institute for Solid State Physics, The University of Tokyo, Kashiwa, Chiba 277-8581, Japan}

\author{Koichiro~Yaji}
\affiliation{Research Center for Advanced Measurement and Characterization, National Institute for Materials Science (NIMS), Tsukuba, Ibaraki 305-0003, Japan}

\author{Shik~Shin}
\affiliation{Office of University Professor, The University of Tokyo, Kashiwa, Chiba 277-8581 Japan}

\author{Fumio~Komori}
\affiliation{Institute for Solid State Physics, The University of Tokyo, Kashiwa, Chiba 277-8581, Japan}

\author{Yohei~Kobayashi}
\affiliation{Institute for Solid State Physics, The University of Tokyo, Kashiwa, Chiba 277-8581, Japan}

\author{Takeshi~Kondo}
\affiliation{Institute for Solid State Physics, The University of Tokyo, Kashiwa, Chiba 277-8581, Japan}
\affiliation{Trans-scale Quantum Science Institute, The University of Tokyo, Tokyo 113-0033, Japan}

\date{\today}


\begin{abstract}
{We use a pump-probe, spin-, and angle-resolved photoemission spectroscopy (ARPES) with a 10.7~eV laser accessible up to the Brillouin zone edge, and reveal for the first time the entire band structure, including the unoccupied side, for the elemental bismuth (Bi) with the spin-polarized surface states. Our data identify Bi as in a strong topological insulator phase ($Z_2$=1) against the prediction of most band calculations. We unveil that the unoccupied topological surface states possess spin-polarized saddle points yielding the van Hove singularity, providing an excellent platform for the future development of opto-spintronics.}
\end{abstract}

\maketitle

Prototype $Z_2$ topological insulators (TIs) are $\rm{Bi_2Se_3}$ and $\rm{Bi_2Te_3}$ for strong TIs~\cite{Zhang2009-TSS_theory, Xia2009-Bi2Se3, Chen2009-Bi2Te3} and $\beta$-$\rm{Bi_4I_4}$~\cite{Liu2016-B_Bi4I4-theory, Autes2016-B_Bi4I4, Noguchi2019-B_Bi4I4} and Bi$_{14}$Rh$_3$I$_9$~\cite{Rasche2013a-Bi14Rh3I9, Rasche2013b-Bi14Rh3I9, Pauly2015-Bi14Rh3I9} for weak TIs, which all contain Bi with strong spin-orbit interaction. Bi is, thus, the most popular element for a material design realizing topological phases. Surprisingly, however, the bulk topology of the elemental Bi itself has not yet been identified and continues to be debated~\cite{Ohtsubo2013-Bi, Ito2016-Bi, Ohtsubo2016-Bi-theory, Fuseya2018-Bi-theory, chang2019-Bi-theory, aguilera2021-Bi-theory, konig2021-Bi-theory}, even though its properties, including spin-polarized surface states, have been vigorously investigated for so many years~\cite{liu1995-Bi-TB, jezequel1997-Bi, ast2001-Bi, koroteev2004-Bi, hirahara2007d-Bi, takayama2011-Bi}. Recently, a possible higher-order topological state has been suggested for the bulk Bi by extending the topological classification to $Z_4$~\cite{Drozdov2014-Bi, Schindler2018-Bi}. However, it is on the basis that Bi is topologically trivial ($Z_2$=0) within the $Z_2$ index. Therefore, identifying the actual topological phase in the elemental Bi is getting more crucial in condensed matter physics.

The bulk topology of Bi can be experimentally determined by identifying how two surface bands (SS1 and SS2) are connected into the bulk conduction and valence bands (BCB and BVB) around \m, as represented in Fig. 1. 
Despite these simple criteria, there are several reasons as follows for having this issue still controversial:

(1)  High-quality Bi is commonly prepared as films. Bi films get free-standing around 15 BLs (6 nm)~\cite{Shirasawa2011-Bi}, and stress from a substrate is totally removed when thicker than $\sim$25 BLs ($\sim$10 nm)~\cite{Yao2016-TI}. However, an interaction between the front and back surface states even in relatively thick films as 200 BLs may open a gap in the surface states and mislead the bulk topology~\cite{Ishida2016-Bi-theory, Fuseya2018-Bi-theory, aguilera2021-Bi-theory}. 

(2) The entire band structure should be determined by experiments for a fair comparison with band calculations. However, this cannot be accomplished by a standard ARPES observing only the occupied states.
One way of visualizing the unoccupied band is to raise the sample temperature and detect thermally excited electronic states above the Fermi level (\e). However, the original bulk topology might be altered by lattice expansion inevitable at high temperatures~\cite{white1972-As_Sb_Bi, monig2005-Bi}. 

(3) The bulk band gap around \m\ is so small ($\sim$15 meV), making it hard to clarify the connections of surface bands to the bulk bands~\cite{tichovolsky1969-Bi, vecchi1974-Bi}. 

In this Letter, we overcome all these difficulties and clarify the genuine topological state in the elemental Bi. 
Following are our solutions (s1)-(s3) to (1)-(3):

(s1) We prepare a film of $\sim$1000 BLs ($\sim$0.4 $\rm{\mu m}$), which is, according to theory~\cite{aguilera2021-Bi-theory}, thick enough to make the overlap of wavefunction between the front and back surface states negligible. 
 
(s2) We use a pump-probe spin-ARPES we recently developed, and unveil the band structure including the unoccupied states over the entire Brillouin zone (BZ). 
Importantly, this technique allows observing unoccupied states without raising the lattice temperature by taking data just after pumping.  

(s3) We employ spin-resolved spectra, which can distinguish between the surface and bulk bands with and without the spin polarization, respectively, to identify whether each of the two surface bands is connected to the conduction band or the valence band.

These experiments conclusively identify the bulk band topology of the elemental Bi to be non-trivial ($Z_2$=1). The state-of-the-art spin-ARPES further reveals a unique feature in the unoccupied surface band: spin-polarized saddle points that form a hexagonal helical spin texture and generate the van Hove singularity (vHS) in the density of states. This could be an iconic structure for the future opto-spintronics application with Bi, which controls the spin current by photoexcitation~\cite{ganichev2002-rev, Yuan2014-WSe2, Jungfleisch2018-Bi, Zhou2018-Bi, hirose2018-Bi}. 

\begin{figure}[t]
  \includegraphics[width=3.3in]{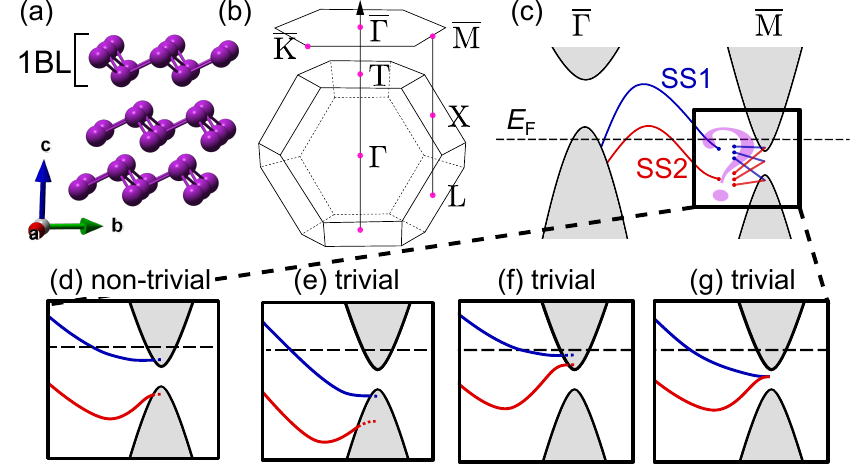}
  \caption{(a) Crystal structure of Bi in the (111) orientation. (b) Brillouin zone for bulk and (111) surface. (c) Schematic band structure along \g\ $-$ \m\ direction on the Bi(111) surface. Blue and red lines show the surface bands (SS1 and SS2) with in-plane spin polarization in opposite directions. (d)-(g) All possible relationships between surface and bulk bands around \m\ corresponding to different bulk topologies.~\cite{Ito2016-Bi}}
\end{figure}

\begin{figure}[t]
  \includegraphics[width=3.3in]{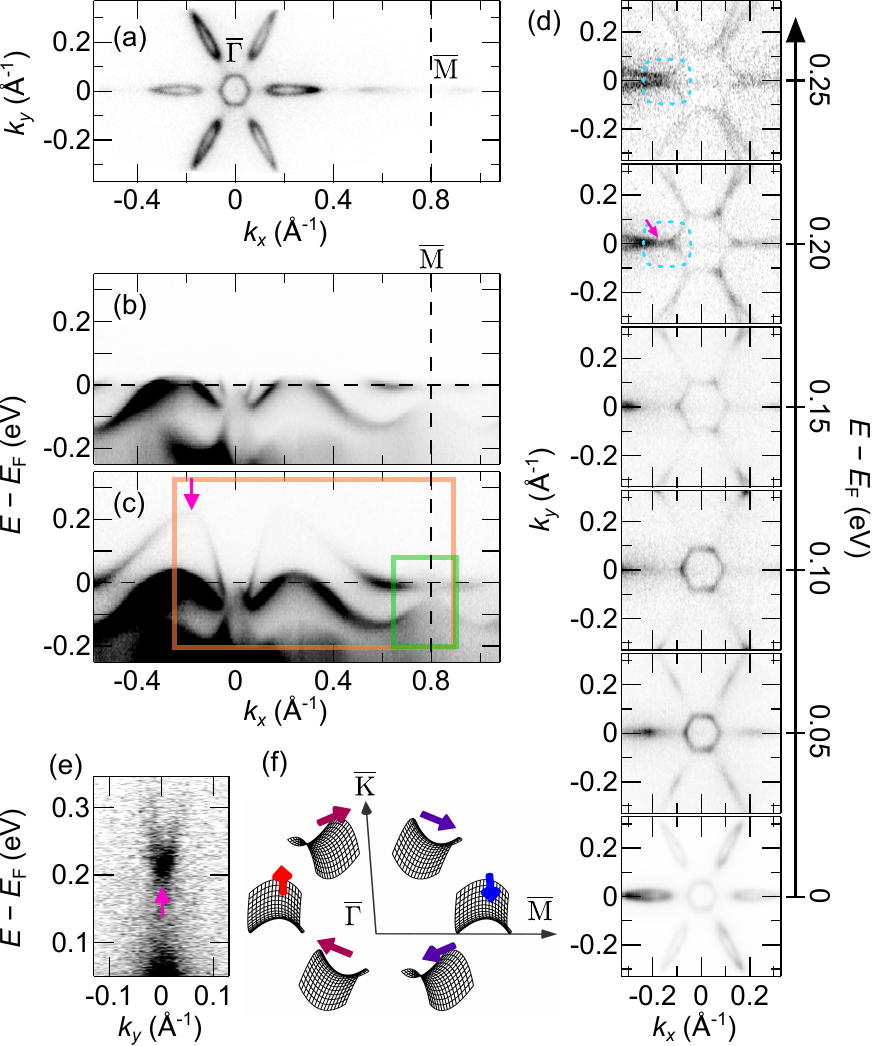}
  \caption{Band structures of Bi revealed by pump-probe ARPES. 
  (a) Fermi surface map. (b,c) Band dispersion along \g\ $-$ \m\ measured without and with pump.  (d) Energy contour maps on the unoccupied side from 0 to 0.25 eV. (e) Band dispersion crossing the saddle point along $k_y$.  (f) Schematic of spin-polarized saddle points with a hexagonal structure. Magenta arrows in (c,d,e) indicate the saddle point.}
\end{figure}

Single-crystal Bi(111) films of $\sim$1000 BL ($\sim$0.4 $\rm{\mu m}$) were prepared $in$ $situ$ by depositing Bi on an Si(111) $7\times 7$ surface (see more details in supplementally materials).  All ARPES measurements were performed with a 10.7 eV laser generated by a home-built Yb:fiber pulse laser ~\cite{zhao2015-laser,zhao2017-laser}. The fundamental Yb:fiber laser (1.19 eV) was also used as a pump light. The energy resolution was $\sim$20 meV and $\sim$25 meV for pump-probe  ARPES and  pump-probe SARPES, respectively. The time resolution was 360 fs. We used a mild pump (0.08 $\rm{mJ/cm^2}$) preventing a lattice vibration. The high repetition rate of the laser (1MHz) and a high-efficiency spin-detector (VLEED) enabled us to obtain a sufficient count rate of spin signals. All experiments were performed around 70 K using $p$-polarized light for both the pump and probe. Details about our newly developed ARPES system will be discussed elsewhere~\cite{Kawaguchi_2023}.


First, we investigate the spin-integrated band structure of Bi(111). Figures 2(a) and 2(b) plot the Fermi surface map and the occupied band dispersion along \g\ $-$ \m\ measured without pumping. We confirm well-known surface bands: a hexagonal electron pocket around \g, petal-like hole pockets surrounding it, and an elongated electron pocket around \m\ ~\cite{ast2001-Bi,koroteev2004-Bi}. These are formed by two surface bands (SS1 and SS2) connecting to bulk bands around \g\ and \m. We further perform the pump-probe measurements and successfully visualize the bands up to the unoccupied side [Fig. 2(c)]. 

Figure 2(d) displays the contour energy maps at different binding energies on the unoccupied side.
Interestingly, we find that hexagonal and petal-like pockets, which are detached from each other around ${E-E_{\rm F}}$~=~0.15~eV, get closer with increasing binding energy.
They touch with each other around 0.2 eV (pointed by a magenta arrow) and eventually turn to continuous parallel segments of an enlarged energy contour (see the regions of dotted light blue rectangles). The touching point locates at the top of the upward energy dispersion along \g\ $-$ \m\ [arrow in Fig. 2(a)] as well as at the bottom of the downward energy dispersion perpendicular to the \g\ $-$ \m\ cut as exhibited in Fig. 2(e). Therefore, this energy state [arrows in Figs. 2(a), 2(d), and 2(e)] is a saddle point, which forms the van Hove singularity in the density of states \cite{van1953-vHS, Du2016-Bi}. The same saddle point is placed at six locations in the surface BZ  [Fig. 2(f)], which forms a helical spin structure, as revealed below. 

The spin polarization is investigated by pump-probe spin-ARPES. Figure 3(b) plots the in-plane $Y$ component of spin polarization ($S_Y$) for the orange rectangular region in Fig. 2(a). Thanks to the pump-probe technique, the spin-polarized states of surface bands (SS1 and SS2) are unveiled not only on the occupied side but also on the unoccupied side. Notably, the sign reversal of spin between SS1 and SS2 is clearly exhibited. Importantly, our experiments demonstrate that the saddle points yielding the vHS around 0.2 eV are spin-polarized with a helical spin structure, as illustrated in Fig. 2(f). This further implies that massive spin currents can be controlled by photoexciting these states with circularly polarized mid-infrared light.

We estimate $|S_Y|$ for the upper and lower surface bands along  \g\ - \m\  in Fig. 3(a) to examine how the spin-polarized surface states are mixed with or absorbed into the bulk states without spin-polarization. The spin polarization should be observed as 100$\%$ for the surface states if the following two conditions are fulfilled~\cite{yaji2017-Bi}. One is that the $E$-$k$ points are far from the time-reversal invariant momenta (\g\ and \m) at which the up and down spins inevitably degenerate. Second is that they are free from hybridization with the bulk states which reduces spin-polarization. As expected, while $S_Y$'s of SS1 and SS2 are close to 100$\%$ in the momentum range far from \g\ and \m\ (0.2 \AA$^{-1}$$<k_x<$ 0.6 \AA$^{-1}$), these decrease and eventually become almost zero at  \g\ and \m. Nevertheless, we find a clear difference between $S_Y(k_x)$'s of the upper and lower surface bands: the latter decreases more rapidly than the former with approaching \g\ and \m\ where the valence bands are situated, as represented by blue arrows in Fig. 3(a). This indicates that the lower surface band is absorbed to (or hybridized with) the balk bands extensively around \g\ and \m. 

\begin{figure}[t]
  \includegraphics[width=3.0in]{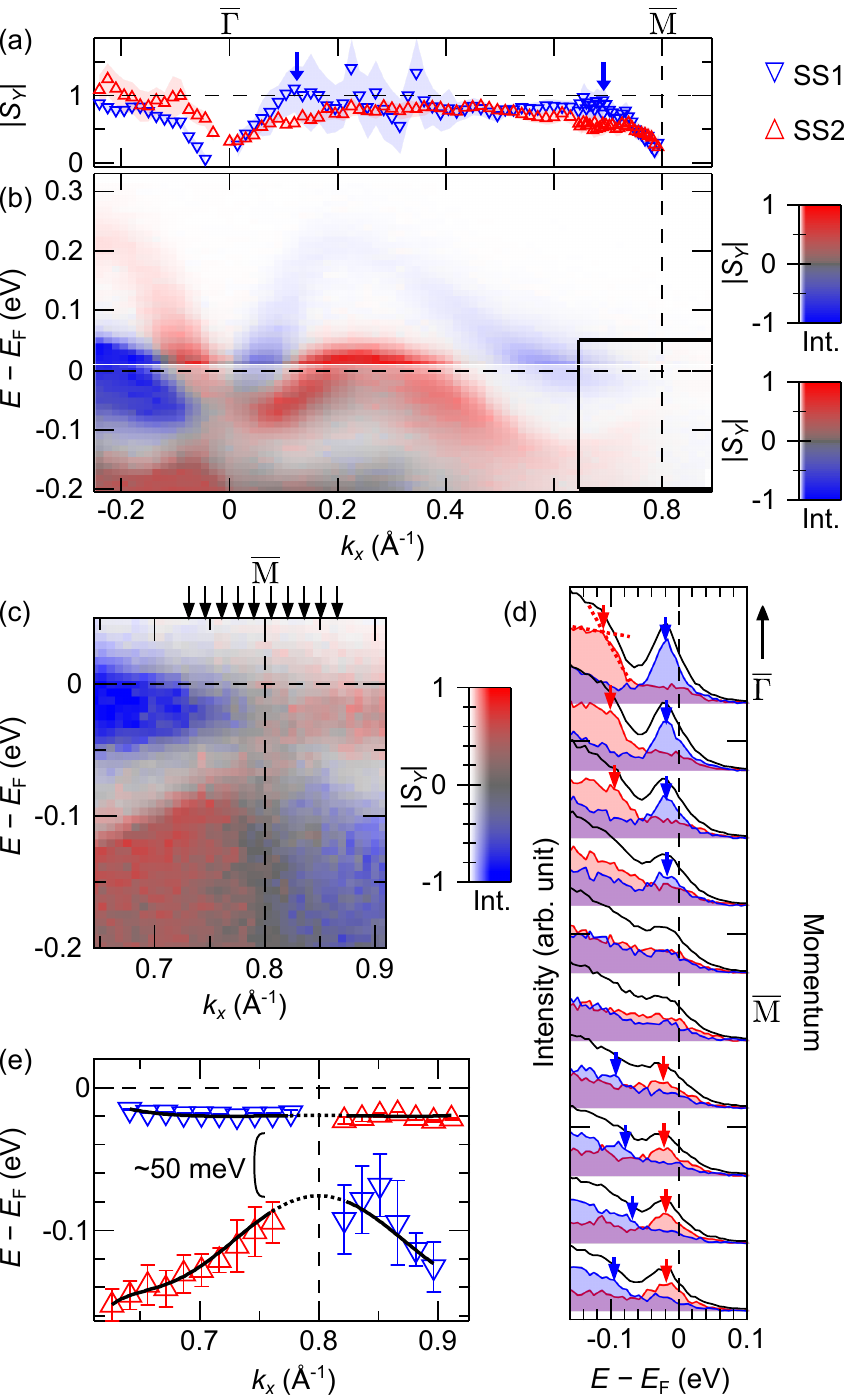}
  \caption{Spin texture of surface states in Bi revealed by pump-probe spin-ARPES. (a,b) Spin polarization and spin-polarized band along \g\ $-$ \m, respectively. Red and blue represent up- and down-spin in the $Y$ direction
for two surface bands (SS1 and SS2). The painted areas represent errors for plots in (a).
(c) High-resolution map of spin polarization around \m\ within the black frame in (b). (d) Spin-resolved EDCs at $k$'s marked by arrows in (c). Energy positions of surface bands are pointed by red and blue arrows. 
Black curves are the addition of the up- and down-spin spectra.
(e) Surface bands determined from spin-resolved EDCs. Fitting curves to the data (solid and dotted lines) are overlayed.
}
\end{figure}

Since the spin-polarization signals originate from the surface states, the surface bands can be determined separately from the bulk states by tracing the peak positions of the spin-polarized spectra. In particular, we measured the spin-polarization map with high precision for the bands around \m\ [Fig. 3(c)], which is the key momentum region to determining the bulk topology of Bi. Figure 3(d) plots spin-resolved energy distribution curves (EDCs) at $k$'s marked by arrows in Fig. 3(c). The spin-integrated EDCs (black lines) are also superposed.  Although peaks are observed for the upper surface band slightly below \e, only a hump structure, poorly defined as a quasiparticle, is obtained for the lower surface band around $-0.1$ eV. This agrees with our assertion that the lower surface band is significantly hybridized with the bulk valence band around \m\ with a broad spectral continuum; the bulk state is observed as a continuum projected onto the surface due to the $k_z$ broadening typical for ARPES, which is a surface-sensitive technique.

\begin{figure}[t]
  \includegraphics[width=3.0in]{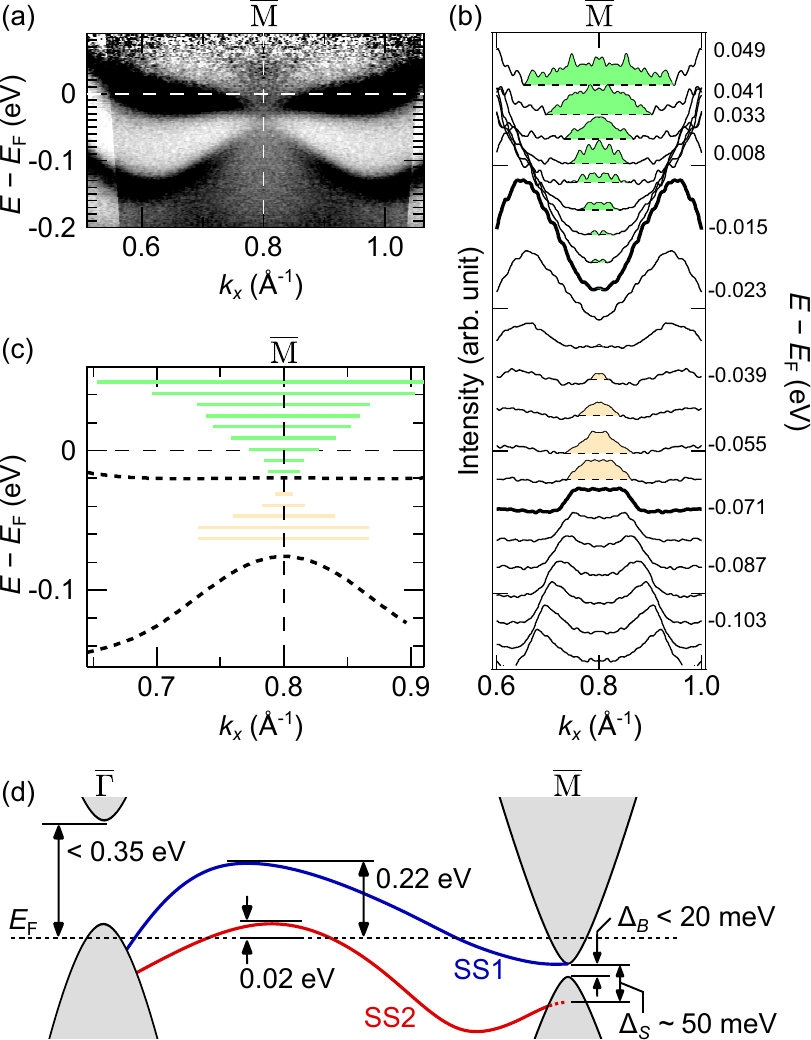}
  \caption{Bulk bands and their relationship with surface bands.
(a) Pump-probe ARPES map around \m. Here, the original spectra are symmetrized across \m\ to remove the matrix element effect. They are also divided by the Fermi-Dirac distribution function at the electron temperature (250 K) estimated from the spectral edge broadening due to the pumping.  
(b) MDCs of (a). Thick black lines represent energies of two surface bands.
Intensities of BCB and BVB are each painted green and orange. 
 (c) Bulk bands (color bars) obtained from (b) and surface bands (dashed lines) determined in Fig. 3(e) are superimposed. 
 (d) Schematic band structure and characteristic energies obtained in our experiments.}
\end{figure}

The spectral hump has a shoulder structure as a remnant of the surface band. The lower surface band is determined by tracing their energies, obtained as the crossing point of two lines fitted to a spectrum, as demonstrated in Fig. 3(d).
 In Fig. 3(e), we plot the results together with the upper surface band. In both bands, plots are missing close to \m, where the spin-polarization is zero; those are, instead, estimated by extrapolating a curve fitted to the data (dotted curves). The upper surface band is almost flat, whereas the lower surface band disperses upward. However, they stay off each other, opening a band gap of $\sim$50 meV at \m. These observations eliminate the case of Fig. 1(g) predicted by most band calculations~\cite{chang2019-Bi-theory,aguilera2021-Bi-theory}.

To pin down the relationship between the surface and bulk bands further, we measure high statistics data of pump-probe ARPES around \m\ [Fig. 4(a)]. The obtained intensity map [Fig. 4(a)] shows a parabola-shaped spectral continuum for the bulk valence and conduction bands (BVB and BCB), other than strong intensities for the surface bands with spectral sharp peaks. To examine the bulk states in more detail, we plot momentum distribution curves (MDCs) around the bulk band gap in Fig. 4(b), where  the intensities for the bulk signals are painted by colors (green and orange for BCB and BVB, respectively). Their intensities reduce the momentum width with approaching each other and eventually disappear without merging together. This indicates that a gap ($<$ 0.02 eV) much smaller than that of the surface bands ($\sim0.05$ eV) opens around $E-E_F=-0.025$ eV. The value of the bulk band gap we observed is consistent with those ($11\sim15$ meV) that have been determined by electromagnetic experiments over the past half-century~\cite{Brown1963-Bi,Smith1964-Bi,Tichovolsky1969-BiSb,Maltz1970-Bi,Vecchi1974-Bi2}. The bulk states are expressed in Fig. 4(c) with color bars. In the same panel, we overlay the upper and lower surface bands (dashed lines) determined from spin-polarized spectra in Fig. 3(e). The result shows that a small portion of the upper surface is absorbed into the bottom of BCB, whereas a large portion of the lower surface band is into BCB around \m, as depicted in Fig. 4(d). The relation between surface and bulk bands corresponds to the case of Fig. 1(a). Hence, the bulk topology of the elemental Bi is non-trivial ($Z_2$=1; strong topological insulator phase), against most theoretical predictions~\cite{chang2019-Bi-theory, aguilera2021-Bi-theory} except for some exceptions~\cite{konig2021-Bi-theory}.

\begin{table}[t]
  \centering
  \caption{Comparison of characteristic energies in band structures obtained: bottom of BCB at \g, top of SS1 and SS2, and energy gap at \m\ between BCB and BVB ($\Delta_{\rm B}$) and between SS1 and SS2 ($\Delta_{\rm S}$). Bulk topology is also listed.}
  \renewcommand{\arraystretch}{1.2}
  \begin{tabular}{c|ccccc|c}
    \hline\hline
    (eV) & BCB (\g) & SS1 & SS2 & $\Delta_{\rm B}$(\m) & $\Delta_{\rm S}$(\m) & $Topology$\\
    \hline\hline
    Our exp. & $<$ 0.35 & 0.22 & 0.02 & $<$ 0.02 & $\sim$0.05 & $non$-$trivial$\\
    Our calc. & 0.21 & 0.18 & 0.04 & 0.06 & 0 & $trivial$ \\
    GGA~\cite{chang2019-Bi-theory}  & 0.29   & 0.26 & 0.06 & 0.1  & 0  & $trivial$\\
    QSGW~\cite{aguilera2021-Bi-theory} & 0.49   & 0.38 & 0.19  & 0.01 & 0 & $trivial$\\
    \hline\hline
  \end{tabular}
\end{table}

In previous studies, the comparison between data and calculations on the bands of Bi has been limited to the occupied states~\cite{liu1995-Bi-TB, aguilera2015-Bi-theory}. In addition, a standard ARPES is out of reach for decisively distinguishing surface and balk bands. These factors have prevented one from fairly evaluating the reliability of band calculations for the elemental Bi. Among modern experimental techniques, a pump-probe spin-ARPES is the only means allowing a full comparison between the data and calculations, and it was first employed for Bi in this work. 

The characteristic values of bands obtained using this unique experimental technique are described in Fig. 4(d): the bottom energy of BCB at \g\ (see details in supplemental materials), the top energies of two surface states (SS1 and SS2), and the energy gaps between BCB and BVB ($\Delta_{\rm B}$) and between SS1 and SS2 ($\Delta_{\rm S}$) at \m. These are compared in Table 1 with the values of the generalized gradient approximation (GGA) and quasi-particle self-consistent GW (QSGW) calculations~\cite{chang2019-Bi-theory,aguilera2021-Bi-theory}. The GGA calculations show good agreement with the data for the bottom of BCB and tops of the two surface bands; however, $\Delta_{\rm B}$(\m) is over-estimated by $\sim$0.1 eV. In contrast, $\Delta_{\rm B}$(\m) shows good agreement in the QSGW calculations, in which, however, the energy position of BCB and two surface bands have large discrepancies of more than 0.1 eV. We also conducted calculations with the Becke-Johnson (BJ) potential, as listed in Table 1. Again, a mismatch with data by more than 0.1 eV is inevitable, and tuning parameters is necessary to obtain a non-trivial phase (see Supplemental Material). 

The debate on the bulk topology of Bi depends on a slight difference of band position in the energy scale merely of several tens of meV. Considering that, the discrepancy as large as $\sim$0.1 eV from experiments revealed here is quite serious for calculations~\cite{konig2021-Bi-theory}. In particular, the unoccupied states unveiled in this work more clearly reveal discrepancies among calculations, providing a strong restriction for evaluating the reliability of the calculations.
Our experimental results not only challenge modern band calculations but also provide good guidelines for the future development of calculational methods to agree with the present experiments eventually.

In conclusion, we revealed the topological nature of the elemental Bi by solving the previous difficulties. 
These experiments consistently reached the conclusion that Bi is in the strong topological insulator phase ($Z_2$=1). Observation of the topological surface states further unveiled fascinating features with the spin-polarized saddle points generating van Hove singularity in the unoccupied density of states. These topological surface bands form a spin helical structure at $\sim$0.2 eV, allowing one to control massive spin current by direct excitation with circularly polarized mid-infrared light. Our results not only challenge theoretical calculations predicting Bi to be a trivial semimetal or a higher-order topological insulator but also indicate that the elemental Bi with a topological nature provides an ideal platform for developing opto-spintronics expected as future technology. 


This work was supported by the JSPS KAKENHI (Grant No. JP21H04439), by MEXT Q-LEAP (Grant No. JPMXS0118068681), and by MEXT as “Program for Promoting Researches on the Supercomputer Fugaku” (Basic Science for Emergence and Functionality in Quantum Matter Innovative Strongly-Correlated Electron Science by Integration of “Fugaku” and Frontier Experiments, JPMXP1020200104) (Project ID: hp200132/hp210163/hp220166).


\end{document}